# Linearly Dichroic Plasmonic Lens and Hetero-Chiral Structures

*Grisha Spektor\*, Asaf David, Bergin Gjonaj, Lior Gal and Meir Orenstein*

Dept. of Electrical Engineering, Technion-Israel Institute of Technology, 32000 Haifa, Israel

\*spektorg@tx.technion.ac.il

**Abstract:** We present theoretical and experimental study of plasmonic Hetero-Chiral structures, comprised of constituents with opposite chirality. We devise, simulate and experimentally demonstrate different schemes featuring selective surface plasmon polariton focusing of orthogonal polarization states and standing plasmonic 'vortex fields'.

**Introduction**

Surface plasmons polaritons (SPPs) are surface electromagnetic waves coupled with collective oscillation of electrons at metal-dielectric interfaces. In the regime below the SPP frequency, where SPP propagation length is substantial, the dominant polarization component of a SPP field propagating along a metal-dielectric interface inside the dielectric is the out-of-plane, longitudinal, component [1, 2]. Therefore when discussing focusing applications we are primarily interested in focusing these fields, namely to generate constructive interference of the longitudinal components in the focal point on the surface. In addition to the higher energy contents, longitudinal field component is desirable because of matching the absorption selection rules of many quantum-well based detectors [3] and can be a source of novel sensing applications and functionality.

Planar chiral symmetry is ascribed to objects lacking in-plane mirror symmetry. When it comes to electromagnetism, it is well known that chiral structures exhibit circular dichroism, that is, they can discriminate the handedness of circularly polarized light. This property was studied, applied and demonstrated in various metamaterials [4, 5]. Finally, another known and rather mystified, fact is that the vast majority of natural materials are *Homo-chiral*, that is, made of constituents of the same chirality.

Probably the most common chiral structure, after the human hand, is the spiral. From the plasmonic perspective, Archimedes' spiral Plasmonic Lenses (SPLs) were proposed, characterized and extensively investigated [6, 7]. It is widely accepted that a SPL with geometrical charge $m = 1$ and a given handedness, distinguishes between right (RHC) and left (LHC) circularly polarized illumination. When illuminated by RHC, a right handed spiral (RHS) will focus the SPPs in the center of the lens and produce a dark spot for the LHC illumination and vice versa. Higher order spirals were shown to produce plasmonic vortices [8, 9, 10]. These properties can be attributed to "spin-orbit" like coupling of the polarization of light and the geometrical spiral phase of the slit. In contrast, much lesser emphasis had been made on linearly dichroic devices. In our work we employed various nano-engraved plasmonic structures in gold comprising different spiral types and their engagements to generate linearly dichroic focusing and higher order caustics of polarized longitudinal fields.

**Work Description**

We study Hetero-Chiral structures resulting from combinations of constituents with opposite chirality. We concentrate on the study of spirals as the chiral building blocks. We present several features of this structural family and discuss the stemming insights.

We show that a combination of two Archimedes' SPLs with opposite handedness (chirality) and unity geometrical charge results in a *Hetero-Chiral* plasmonic lens (HC lens) (Fig. 1a schematics, Fig. 2a – the actual device and Fig. 3 results) with a linearly dichroic focal spot of the longitudinal field. In such merging, the chiral symmetry of the constituents is broken, however the mirror symmetry is not completely restored by this construction, implying that the combined structure should distinguish between horizontal and vertical excitations leading to linear dichroism. The simulation results of this structure, presented in Fig. 3, exhibit high contrast linear dichroism.

A complementary view of the behavior of the structure is through the superposition principle. We can express the linear polarization basis elements in terms of the circular polarization basis elements. In essence this would result in $\hat{x} \propto RHC + LHC$ and $\hat{y} \propto RHC - LHC$ with an imaginary proportionality constant. As stated above, each constituent spiral equally focuses the circular component with the matching handedness. By construction, the field in the focal spot of the HC lens is the linear superposition of the focal spot fields of the constituent spirals. When the HC lens is illuminated by $\hat{x}$ polarization, the focal spot fields add up, resulting in a constructive SPP focus. Conversely, when illuminated with the orthogonal $\hat{y}$ polarization the constituent contributions cancel out, resulting in a dark focal spot.

A further simplification of the 2 engaged spirals can be gained by removing the central part of the structure (e.g. Fig. 1b) resulting in a Hollow HC lens. For given confining dimensions, depending on the geometrical charge of the constituent chiral block, the total structure will resemble a distorted circle. Intriguingly, this slight deformation from perfect circle results in a strong linear dichroism at the focal spot (Fig 4). As we increase the charge of the constituent spirals the extinction ratio between the two polarizations increases.

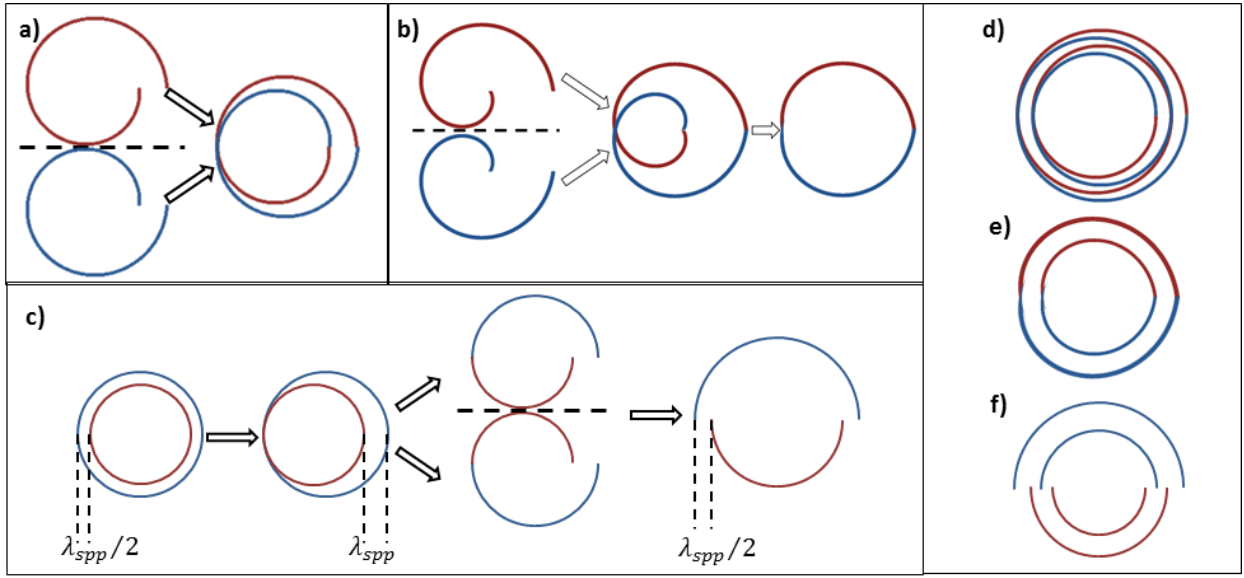

Figure 1: (a) Two chiral constitutes form an achiral combination. The black dotted line is the mirror "asymmetry" plane. (b) Removal of the core of the engaged spirals, Hollow HC lens with constituent spirals of geometrical charge $m = 4$. (c) Formation of the 'Half-Moons' lens by the HC logic (d-f) Schematic demonstration of resonant enhancement of all the HC lenses. The red and blue colors assist in tracking the parts of the constituent elements.

Proceeding with the resemblance to circles and expanding the Hetero-Chiral approach, we examine Fig. 1c. Starting with a non-chiral structure comprised of two concentric circles with $\lambda_{spp}/2$ radii difference, we shift the inner circle by the radii difference with respect to the outer circle, and we obtain a structure similar to the HC lens in Fig. 1a. The latter can be easily disengaged to exhibit its hetero-chiral nature – now comprised of a pair of non-Archimedean spirals. Each is constructed by shifted half circles with increasing radii. These, Half-Circle based spirals are interesting in their own right but are outside the scope/focus of this paper. They lack the geometrical phase of the Archimedes spirals and, among other features, are expected to outperform them in terms of focusing SPPs produced by linearly polarized illumination. As a further step we can shift the small arc with respect to the large one resulting in a structure consisting of two concentric, angularly non-overlapping arcs with radii difference of $\lambda_{spp}/2$. This Half-Moon lens has the same symmetry properties as the hetero-chiral structure presented above. It is mirror-symmetric in the vertical plane and mirror-asymmetric in the horizontal plane, capable of linear dichroism by the same symmetry arguments as above. The performance of this lens is shown in Fig. 5 exhibiting strong linear dichroism of the focal spot. A similar structure was studied by [11].

The focal spot field intensity can be enhanced by introducing additional slits parallel to the curvature of the initial structures (Fig 1. (d-f)). The SPPs launched by the additional corrugations will constructively interfere provided the slits are spaced a distance equal to an integer multiple of $\lambda_{spp}$. The extension of the basic HC lens is achieved by engaging spirals with multiple arms as the chiral building block. An example is the engaged double arm spirals of Fig. 1d. This approach is limited, of course, by the SPP propagation length.

All our proposed designs were extensively simulated using a commercial simulator based on the FDTD method [12]. The results of the simulations are brought alongside the experimental results in Figs. 3-6.

**Fabrication and Experimental Measurements**

The HC structures were fabricated by FIB patterning of a $175[nm]$ thick gold film over glass substrate. The samples were illuminated by the requested polarization states using a semiconductor laser with $\lambda_0 = 671[nm]$. the thickness of the milled slits was $\sim 0.23\lambda_0$ as suggested by [13] to maximize the SPP coupling efficiency. The transmitted near field pattern was collected using an aperture-less NSOM with a metal tip (measuring mainly the longitudinal fields) assisted by pseudo heterodyne interferometry to measure simultaneously phase and amplitude [14]. The experimental results presented in Figs. 3-6 exhibit excellent match to the simulation results.

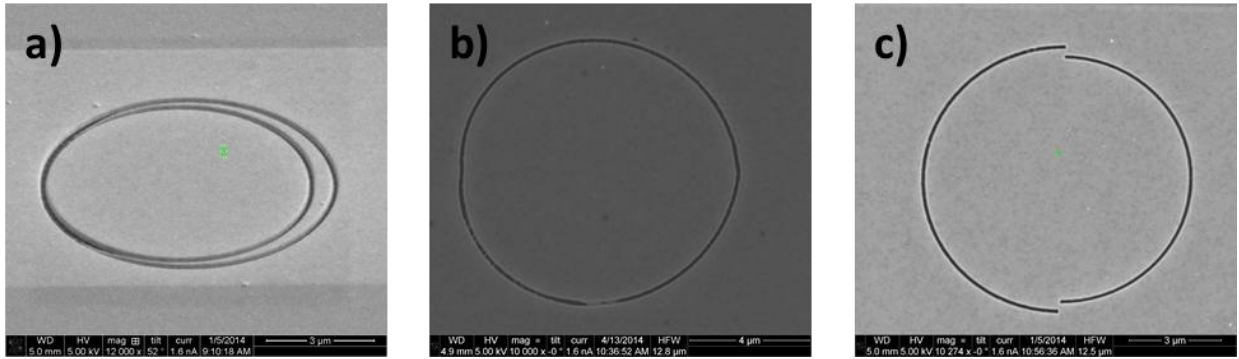

Figure 2. SEM Images of the fabricated lenses. The HC lens (a), Hollow HC Lens (b) and the Half Moons Lens (c).

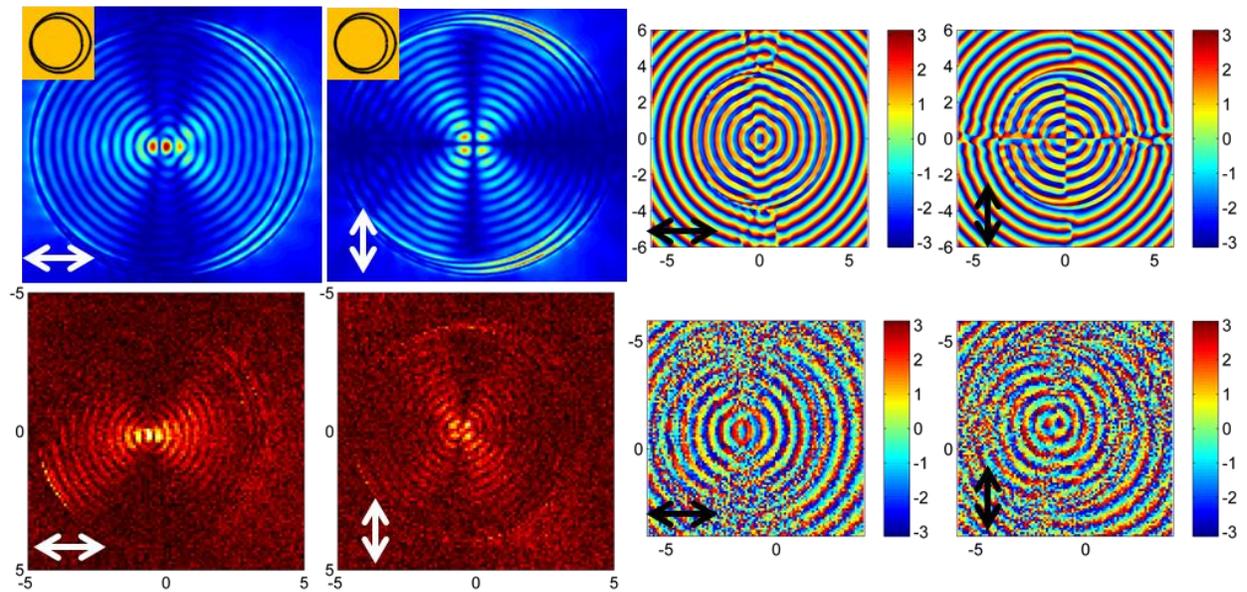

Figure 3. HC Lens. Comparison of FDTD simulation results (Top Row) and experimental NSOM measurements (Bottom Row). The black and white arrows insets represent the polarization of the illuminated light. The 4 images on the left are the amplitude of the SPP field and the images on the right are the phase. The focal spot clearly distinguishes between the orthogonal linear polarizations resulting in constructive focusing of one and destructive focusing of the other.

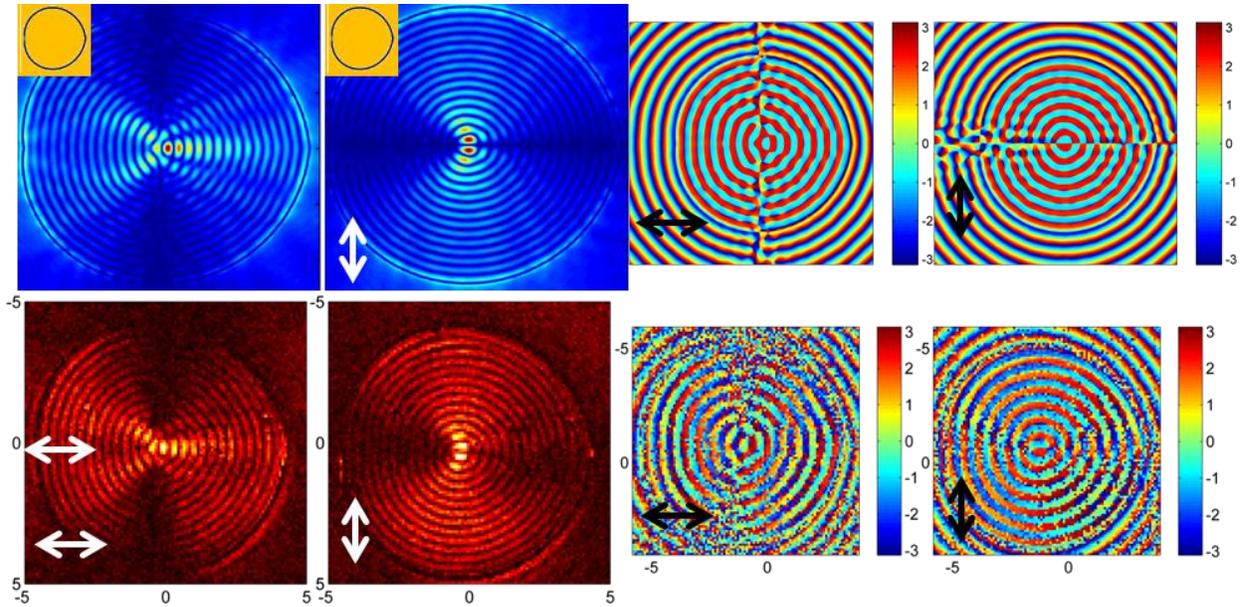

Figure 4. Hollow HC Lens, m=4. Comparison of FDTD simulation results (Top Row) and experimental NSOM measurements (Bottom Row). The black and white arrows insets represent the polarization of the illuminated light. The 4 images on the left are the amplitude of the SPP field and the images on the right are the phase. The focal spot clearly distinguishes between the orthogonal linear polarizations resulting in constructive focusing of one and destructive focusing of the other.

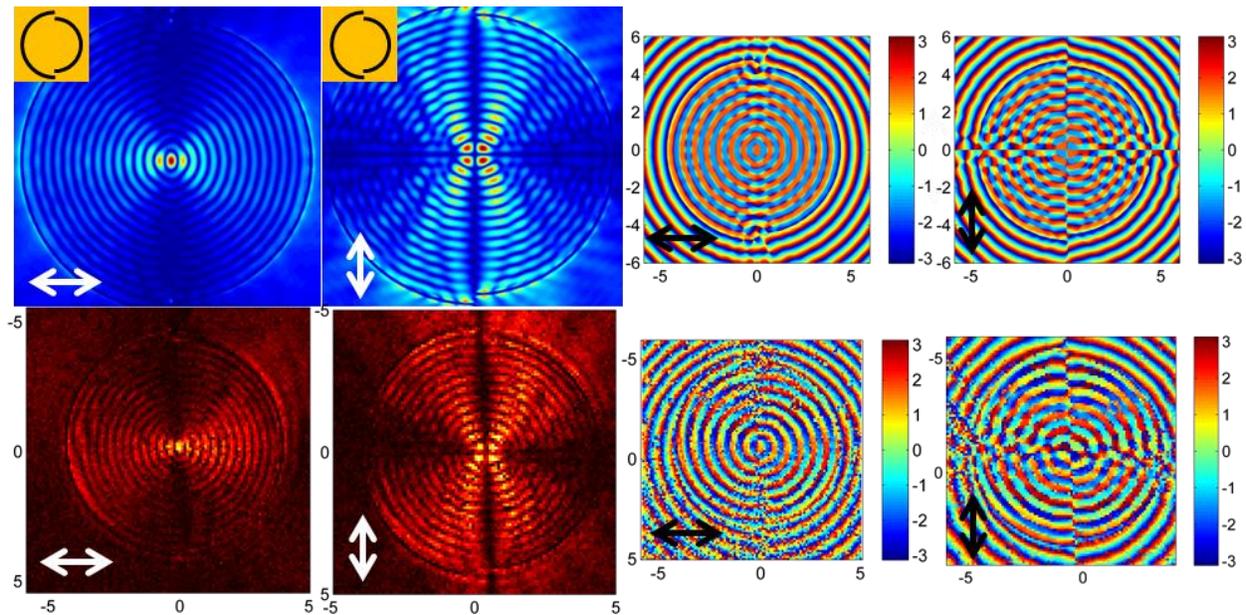

Figure 5. 'Half-Moons' Lens. Comparison of FDTD simulation results (Top Row) and experimental NSOM measurements (Bottom Row). The black and white arrows insets represent the polarization of the illuminated light.

The 4 images on the left are the amplitude of the SPP field and the images on the right are the phase. The focal spot clearly distinguishes between the orthogonal linear polarizations resulting in constructive focusing of one and destructive focusing of the other.

Finally, we constructed higher order Hetero-Chiral structures comprised of higher order spirals for creation of standing vortex waves and focusing to more elaborated caustics than a focal point ( [8, 9, 10] proposed the concept of a Plasmonic Vortex lens PVL). The basic PVL of integer order $m$ consists of discrete pieces of truncated Archimedes spirals of topological charge $m$ placed with azimuthal periodicity equal to the topological charge. The PVL couples the angular momentum of the illumination to SPPs adding an orbital angular momentum of order $m$ resulting in a vortex behavior of the focal spot. When applied to the hetero-chiral vortex lens, we were able to produce a structure which upon illumination with linearly polarized field, is generating projections of both counter-rotating vortices interfering to a pattern with 2m angular intensity lobes (Fig. 6).

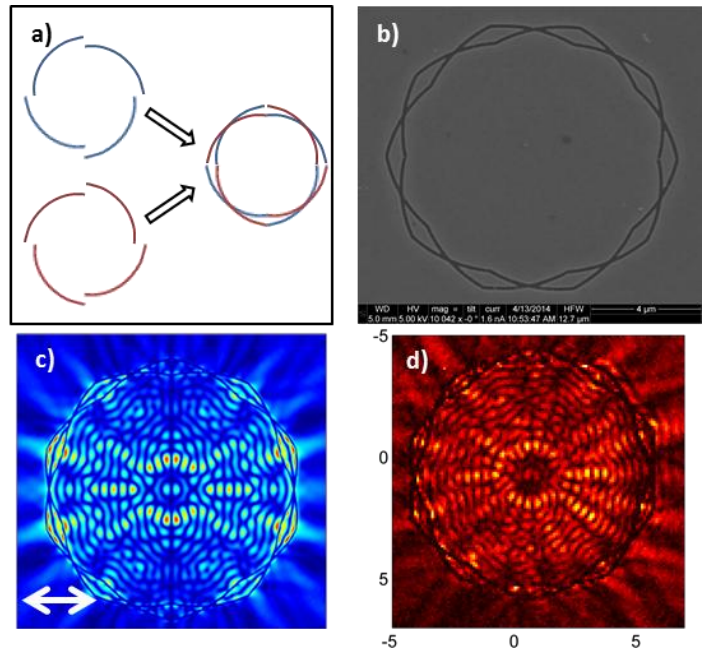

Figure 6. Schematic (a), SEM Image (b), FDTD simulation (c) and NSOM measurement (d) of the azimuthal pattern of order m=10

## Summary and application

Plasmonic Hetero-Chiral structures, comprised of constituents with opposite chirality were studied theoretically and experimentally – showing significant linear dichroism in the focal point.